\newcount\opcion
\newlength{\anchopar}
\def \bfr {\begin{flushright}}
\def \efr {\end{flushright}}
\def \caja {\makebox[3.2cm][1]}

\def \d {\hbox{d}\,}

\def \e {\hbox{e}}

\def \v {\vskip}

\def \p {\partial}

\def \be {\begin{equation}} 
\def \ee {\end{equation}} 

\def \ba {\begin{array}}
\def \ea {\end{array}}

\def \bea {\begin{eqnarray}}
\def \eea {\end{eqnarray}}

\def \bfr {\begin{flushright}}
\def \efr {\end{flushright}}
\def \caja {\makebox[3.2cm][1]}

\def \m {\overline{m}}
\def \f {\overline{f}}

\documentstyle[a4,12pt,psfig]{article}

\ifnum\opcion=1
\else

\fi

\begin{document}

\pagestyle{empty}
\bfr
\caja{{\tt FTUV/94-23}}\\
\caja{{\tt IFIC/94-21}}\\
\caja{{\tt Imperial-TP/93-94/32}}\\
\caja{{\tt hep-th/9405015}}
\efr
\v 1cm

\vfil

\begin{center}

{\Huge Canonical Structure of 2D Black Holes.{\Large\footnote[2]{Work 
partially supported by the C.I.C.Y.T. and the D.G.I.C.Y.T.}}}

\v 0.3cm

Jos\'e Navarro-Salas$^{1,2}$, Miguel Navarro$^{2,3,4}$\\
and C\'esar F. Talavera$^{1,2,3}$ 

\v 0.3cm
\end{center}

\noindent 1.- Departamento  de  F\'\i sica  Te\'orica, Burjassot-46100, 
Valencia, Spain. 

\noindent 2.- IFIC, Centro Mixto Universidad de
Valencia-CSIC, Burjassot-46100, Valencia, Spain.

\noindent 3.- The Blackett Laboratory, Imperial College, London SW7 2BZ, 
United Kingdom.

\noindent 4.- Instituto Carlos I de F\'\i sica Te\'orica y Computacional, 
Facultad  de  Ciencias, Universidad de Granada, Campus de Fuentenueva,
18002, Granada, Spain.


\vfil

\begin{center}{\bf Abstract}
\end{center}
We determine the canonical structure of two-dimensional black-hole solutions 
arising in $2D$ dilaton gravity. 
By choosing the Cauchy surface appropriately we find that the
canonically conjugate variable to the black hole mass is given by the
difference of local (Schwarzschild) time translations at right and left
spatial infinities. 
This can be regarded as a generalization of Birkhoff's theorem.

\vfil
\eject


\setcounter{page}{1}
\pagestyle{plain}

Two-dimensional dilaton gravity models have attracted much 
attention in the last few years because they are useful toy 
models for quantum gravity. The string inspired model proposed by 
Callan, Giddings, Harvey and Strominger (CGHS-model) \cite{[CGHS]} 
(see the review \cite{[Harvey]})  
provides an excellent scenario to study black-hole physics 
(the black hole solutions were first found in 
\cite{[Witten],[Mandal]}). 
An exact, non-perturbative solution to 
the theory is still lacking, although some progress has been 
achieved recently \cite{[Hirano],[Verlinde],[Mikovic]} using canonical 
quantization methods. 

In any canonical quantization approach the issue of the reduced 
phase-space of the theory is of great importance. 
For the CGHS-model this question 
was addressed in ref. \cite{[MikovicandNavarro]} (see also 
\cite{[NavarroandTalavera]}) following the strategy of the covariant phase-space 
formalism \cite{[CrnkovicandWitten],[NPB]}. However only the case of compact 
spatial section was analysed in \cite{[MikovicandNavarro]} owing to the 
subtleties of the non-compact case. A similar study was carried 
out in \cite{[Soh]} although their results are valid for the case 
of compact spatial section only. The aim of this paper is to 
perform a rigorous analysis of the canonical structure of the 
model for an open spatial section. We shall 
restrict ourselves to the case of pure gravity because the 
relevant questions we want to address already emerge in the pure 
gravity situation. We shall also consider spherically 
symmetric Einstein gravity, which is equivalent to a $2D$ dilaton gravity 
model. This model has been analysed in a rather involved way 
through Ashtekar's hamiltonian approach \cite{[Thieman]}.

In the covariant formulation of the canonical formalism the phase 
space is defined as the space of all classical solutions. This 
space is endowed with a presymplectic two-form $\omega$
\be 
\omega = 
\int_\Sigma \d\sigma_\mu(-\delta j^\mu)\>,\label{1}
\ee
where $\Sigma$ is a Cauchy hypersurface and $-\delta j^\mu 
=\omega^\mu$  (the symplectic current) can be obtained from the 
variation of the action $S=S[\Psi^\alpha]$
\be 
\delta S =\int \partial_\mu j^\mu +  
 {\delta S\over \delta \Psi^\alpha}\delta 
\Psi^\alpha  \label{2}\quad.
\ee

Now two remarks are in order. The presymplectic form (\ref{1}) is 
constructed in the same way as a Noether charge. Therefore, 
without suitable boundary conditions (\ref{1}) is not a 
well-defined quantity (finite and independent of $\Sigma$). Moreover, 
the two-form (\ref{1}) is not necessarily non-degenerate. In 
general, it could have a non-trivial kernel. We can define the 
infinitesimal gauge-type symmetries of the theory as those 
generated by the kernel of the presymplectic form $\omega$ in 
(\ref{1}). The physical (non-degenerate) symplectic form is 
obtained by pushing down (\ref{1}) on the quotient manifold with 
respect to the degenerate directions (i.e., the reduced 
phase-space). These two crucial points will be discussed 
throughout the paper. 

Let us start our analysis of the pure gravity CGHS model 
\bea 
S_{CGHS} &=& \frac{1}{2}\int_{\cal M}\,d^2x\sqrt{-g}
\left[\e^{-2\phi}(R+4(\nabla\phi)^2 +4\lambda^2)\right]\>,\label{3}
\eea
by writing down the potential one-form $j^\mu$ for the 
symplectic current. Following the scheme outlined above we obtain 
\bea 
j^\alpha &= &\frac12\sqrt{-g}\e^{-2\Phi}\left(8\p^\alpha\Phi\delta\Phi - 
2\p^\alpha\Phi(g^{\mu\nu}\delta g_{\mu\nu}) - 
2\p_\beta\Phi\delta g^{\alpha\beta}\right.\nonumber\\
&&+ \left. g^{\mu\nu}\delta\Gamma^\alpha_{\mu\nu} 
- g^{\mu\alpha}\delta\Gamma_{\mu\nu}^\nu\right)\label{4}\>.
\eea
The next point is to insert the general solution of the classical 
equations of motion into (\ref{4}). It is well known that, up to 
space-time diffeomorphisms, the general solution of the model is 
given by the black hole solutions. In the conformal gauge, 
$\d s^2 = -\e^{2\rho}\d x^+\>\d x^-\>$, it can be written as  
\be 
\e^{-2\Phi}=\e^{-2\rho}=\frac{m}{\lambda} - \lambda^2x^+ x^-
\>,\label{5}
\ee
where $x^+,\> x^-$ play the role of the null Kruskal coordinates 
of the Schwarzschild black hole and the parameter $m$ turns out 
to be the ADM mass of the black hole \cite{[Witten],[Bilal]}. 
To recover standard static black 
hole solutions one can introduce the local coordinates 
$\tau=\frac12(\sigma^++\sigma^-)$, 
$\sigma=\frac12(\sigma^+-\sigma^-)$, with
\bea 
\lambda x^+ &=& \e^{\lambda \sigma^+}\label{6}\>,\\
\lambda x^- &=& -\e^{-\lambda \sigma^-}\label{7}\>,
\eea
arriving at 
\bea 
\d s^2&=& -\left(1+\m\e^{-2\lambda 
\sigma}\right)^{-1}\d \sigma^+\d\sigma^-\label{8}\>,\\ 
\e^{-2\Phi}&=&\m+\e^{2\lambda\sigma}\label{9}\>,
\eea
where $\m =m/\lambda\>.$
In the asymptotic region, $\sigma\rightarrow+\infty$, the solutions 
approach to the linear dilaton vacuum (LDV)
\be 
\rho\sim0\quad,\Phi\sim-\lambda\sigma\label{10}\>.
\ee

We should remark that, in contrast to the standard Poincar\'e invariance of 
the flat space-time vacuum of general relativity, the only 
symmetry of the LDV is the time translation. In other words, the 
time translation is the only asymptotic symmetry of the black 
hole solution (\ref{8}, \ref{9}). Closely related to this is the 
issue of the gauge character of the space-time diffeomorphisms. At 
an intuitive level one could expect that only those 
diffeomorphisms leading to  trivial Noether charges should 
be regarded as gauge symmetries \cite{[Coleman]}. This is 
in accord with our covariant definition of gauge-type 
symmetries. If the transformation generated by a vector field 
$X_H$ leaves $\omega$ unchanged and $X_H$ belongs to the kernel of 
$\omega$, then the Noether charge $H$  
($\hbox{i}_{X_H}\omega=\delta H$) vanishes on the covariant 
phase space. In general relativity the unique diffeomorphisms 
leading to non-trivial charges are the asymptotic Poincar\'e 
transformations \cite{[Ashtekaretal]} (see also 
\cite{[Crnkovic]}). In our case, and due to the dilaton field, 
the only transformations that preserve the asymptotic behaviour 
of the fields (\ref{10}) are the time translations. 
Therefore, the most general admissible solution 
is given by (\ref{9}) and 
\be 
\d s^2 = (1+{\overline{m}}\e^{-2\lambda\sigma})^{-1}
\left(-\left(\d(\tau+f(\tau,\sigma))\right)^2 + (\d\sigma)^ 2 \right)\>,\label{11}
\ee
where $f(\tau,\sigma)$ is an arbitrary function with vanishing 
derivatives at spatial infinity.

To evaluate $j^\alpha$ on-shell we first need to evaluate the 
Christoffel symbols:
\bea
\Gamma^0_{00}&=&-\frac{mf'(1+\dot f)}{{\overline{m}}+ 
\e^{2\lambda\sigma}} +\frac{\ddot f}{1+\dot f} \>,\\  
\Gamma^0_{01}&=&\frac{m(1- {f'}^2)}{{\overline{m}}+ 
\e^{2\lambda\sigma}} +\frac{\dot {f'}}{1+\dot f}\>,\\  
\Gamma^0_{11}&=&\frac{mf'(1-{f'}^2)}{({\overline{m}}+ 
\e^{2\lambda\sigma})(1+\dot f)} +\frac{f''}{1+\dot f} \>,\\  
\Gamma^1_{00}&=&\frac{m(1+\dot f)^2}{{\overline{m}}+ 
\e^{2\lambda\sigma}} \>,\\  
\Gamma^1_{01}&=&\frac{mf'(1+\dot f)}{{\overline{m}}+ 
\e^{2\lambda\sigma}} \>,\\  
\Gamma^1_{11}&=&\frac{m(1+ {f'}^2)}{{\overline{m}}+ 
\e^{2\lambda\sigma}} \>.
\eea
Pushing down the current (\ref{4}) on the space of solutions 
(\ref{9}), (\ref{11}), and after a long computation, the 
 expression for $j^\mu$ is calculated to be 
\bea 
j^\mu &=&\frac12\epsilon^{\mu\nu}\left[\p_\nu\left(2m\delta f
+({\overline{m}} +\e^{2\lambda\sigma})\delta f'-
(\m+\e^{2\lambda\sigma})\frac{f'}{1+\dot f}\delta \dot 
f\right)\right.\nonumber\\
&&\left.-4\lambda\e^{2\lambda\sigma}\p_\nu\delta 
f\right]\label{18}\>.
\eea

In the static region $-\infty<\sigma^+,\sigma^-<+\infty$ 
the symplectic current turns out to be then 
\bea 
\omega^\mu=-\delta j^\mu&=&
-\frac12\epsilon^{\mu\nu}\p_\nu\left(2\delta m \wedge \delta f + 
\delta\m\wedge\delta f' - \frac{f'}{1+\dot f}\delta \m \wedge \delta \dot 
f\right.\nonumber\\
&&\left.-\frac12(\m+e^{2\lambda\sigma})\frac{1}{1+\dot f}\delta 
f' \wedge \delta \dot f\right)\>.\label{19}
\eea
Therefore the contribution of this asymptotically flat domain to 
the two-form $\omega$ is given by 
\be 
\omega_I = \left. \frac12\left(2\delta f \wedge \delta m + 
\delta f' \wedge \delta \m 
- \frac{f'}{1+\dot f}\delta \dot f \wedge \delta m - \frac{\m + 
\e^{2\lambda\sigma}}{1+\dot f}
\delta\dot f \wedge \delta f'\right) 
\right|_{\p \Sigma_I}\label{20} \>,
\ee
where $\p \Sigma_I$ represents the two spatial ends ($+\infty$ and 
$-\infty$). To find a finite resulting expression for 
(\ref{20}) we have to require appropriate boundary conditions.  
The minimal requirement is to assume 
\be 
\e^{\lambda\sigma} f', \e^{\lambda\sigma} \dot f 
\sim_{_{\!\!\!\!\!\!\!\!\!{\sigma\rightarrow +\infty}}}\> 
0 \label{21} \>,
\ee
and 
\be 
f' \sim_{_{\!\!\!\!\!\!\!\!\!{\sigma\rightarrow 
-\infty}}}\> 0\label{22} \>.
\ee
These conditions ensure the finiteness of (\ref{20}) as well 
as the independence from the particular surface
$\Sigma_I$:
\be
\omega_I = \delta( f(+\infty) - f(-\infty) ) \wedge \delta m \>. \label{22i}
\ee
The above result suggests to choose the Cauchy surface $\Sigma$ in such
a way that it connects the right and left infinities through the
static regions I and II of Kruskal diagram (see Fig. I). Adding to
(\ref{22i}) the contribution of the asymptotically flat region II we have
\be 
\omega = \omega_I+\omega_{II}= 
\delta \left(f(i^0_R)-f(i^0_L)\right)\> \wedge \delta 
m\>,\label{23}
\ee
where $i^0_R$, $i^0_L$ stands for the right and left spatial
infinities. 
This means that the reduced phase-space  of the model is 
two-dimensional, where the canonically conjugate variable to the mass 
is the global variable 
$\left(f(i^0_R)-f(i^0_L)\right)\equiv \f$ 
of the local time translations. Observe that the parameter 
$m$ clearly emerges now as the Noether charge of the asymptotic 
time translation $\left(\f\rightarrow \f + a^0\right)$:
\be
\hbox{i}_{({\frac{\p\>}{\p\f}})}\>\omega = \delta 
m\>.\label{24}
\ee

\ifnum\opcion=1
	\begin{figure}
	\centerline{\psfig{figure=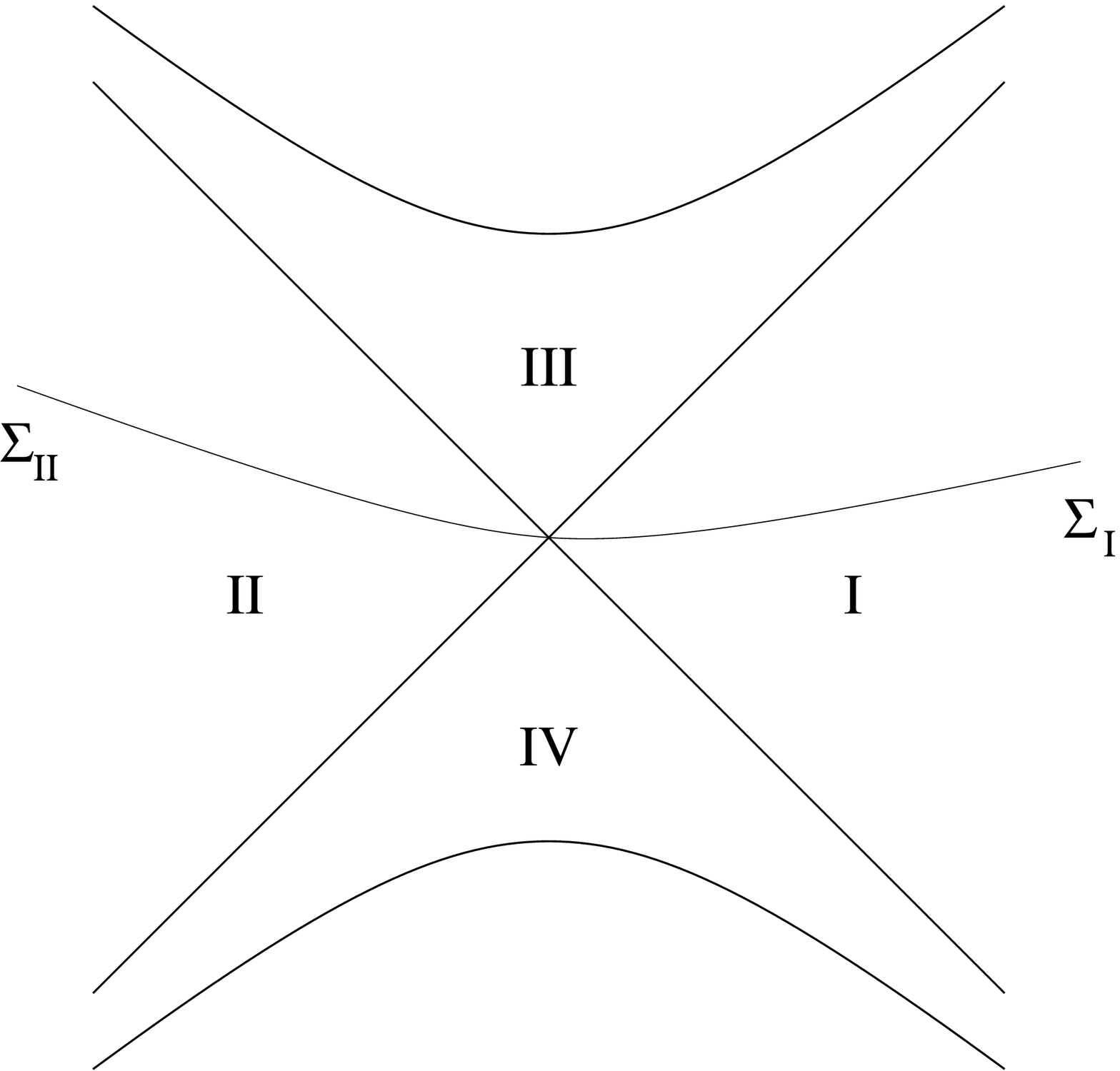,width=4in}}
	\caption{Kruskal diagram for black-hole spacetime. 
		$\Sigma=\Sigma_I \cup \Sigma_{II}$ is the Cauchy surface.}
	\end{figure}
\else
\fi

Now we want to extend our previous discussion to the case of
spherically symmetric Einstein gravity. Although this theory can be
seem as a special case of a general class of two-dimensional dilaton
gravity models, we shall treat it as a four-dimensional theory. 

The symplectic current potential of general relativity is 
\be 
j^\alpha = 
{1 \over 16\pi } \sqrt{g}\left( g^{\mu\nu}\delta \Gamma^\alpha_{\mu\nu}- 
g^{\mu\alpha}\delta \Gamma^\nu_{\mu\nu}\right)\>.\label{25}
\ee

According with Birkhoff's theorem \cite{[Birkhoff]} the Schwarzschild 
solution represents the general solution, up to space-time
diffeomorphisms, of spherically symmetric gravity in vacuum. To work
out the canonical structure of the model one can proceed as in the
previous case. Taking into account that the only surviving asymptotic
Poincar\'e  symmetry of the reduced  
theory is the time translation, the most general admissible solution
in region I and II has to be of the form
\bea
\d s^2&=&- (1-\frac{2m}r) \d^2(t+f(t,r))+(1-\frac{2m}r)^{-1}\d r^2 \nonumber\\
&&+r^2(\d \theta^2+\sin^2\theta\d \varphi^2)
\>. \label{26}
\eea
Inserting (\ref{26}) into the time component of the current (\ref{25})
we obtain 
\be 
j^0 = {1 \over 16\pi } \sin\theta\frac{\d}{\d r}\left[-\frac{r(r-2m)}{1+\dot
f}f'\delta \dot f + r(r-2m)\delta f'- 2f\delta m + 2m\delta f\right]
\>. \label{27}
\ee
Integrating over $2 m < r < +\infty$, $0< \varphi < 2 \pi$, $0 < \theta <
\pi$, it is now easy to arrive at the following expression for $\omega_I$
\be 
\omega_I = {1 \over 4 } 
\left.\left[-2\frac{rf'}{1+\dot f}\delta m\delta \dot f +
\frac{r(r-2m)}{1+\dot f}\delta f'\delta \dot f + 2r\delta m\delta f' +
4\delta f\delta m\right]\right|^{r=+\infty}_{r=2m}
\>. 
\label{28}
\ee 
For $\omega_I$ to be well defined the asymptotic behaviour
has to be adjusted appropriately. We can require the following fall-off
behaviour 
\be 
r f', r \dot f \>\sim_{_{\!\!\!\!\!\!\!\!r\rightarrow\infty}}\> 
0 
\>, \label{29}
\ee
and
\be
f' \> \sim_{_{\!\!\!\!\!\!\!\!r\rightarrow 2m}}\> 0 \>. \label{29i}
\ee

Assuming (\ref{29}), (\ref{29i}), we arrive at 
\be 
\omega_I = \delta \left(f(+\infty)-f(2m)\right) \wedge \delta m
\>,\label{30}
\ee
and the final symplectic form is
\be
\omega = \delta( f(i^0_R) - f(i^0_L) ) \wedge \delta m \> .
\label{30bis}
\ee
We can observe the parallelism between the results (\ref{30bis}) and
(\ref{23}). It seems to be a general property of black hole solutions
and constitutes a kind of generalization of Birkhoff's theorem in the
sense that identifies the true dynamical degrees of freedom.
As a byproduct, this implies that the conformal gauge does 
not capture the full canonical content of the two-dimensional 
metric in a non-compact spatial 
world. In going to the conformal gauge 
one makes use of some particular  diffeomorphisms 
that cannot be regarded as gauge transformations.

	The explicit knowledge of the reduced phase space of
lagrangian models opens an avenue for their quantization 
\cite{[Wittenii],[NPB]} 
and, at the same time, could suggest new variables for quantizing the
system at a presymplectic level. We plan to address these issues in a
future publication extending the present work to the case when matter
fields are present.

        After completing this work we received a preprint \cite{[kuchar]}
where the canonical structure of Schwarzschild black holes is analyzed with
different methods. Our result for spherically symmetric gravity agrees with
that of Ref. \cite{[kuchar]}.

\section*{Acknowledgements} M. Navarro acknowledges to the spanish MEC for a 
Postdoctoral fellowship. C.F. Talavera is grateful to the 
{\it Generalitat Valenciana} for a FPI grant.

\ifnum\opcion=1
\else
	\setlength{\anchopar}{\hsize-\rightskip-\leftskip}
	\newpage

	\pagestyle{empty}
	\section*{Figure Captions}
	Figure I: Kruskal diagram for black-hole spacetime.
	$\Sigma = \Sigma_I \cup \Sigma_{II}$ is the Cauchy surface.

	\newpage
	\pagestyle{empty}
	\centerline{\Large FIGURE I}
	\vfill
	\begin{figure}[h]
	\centerline{\psfig{figure=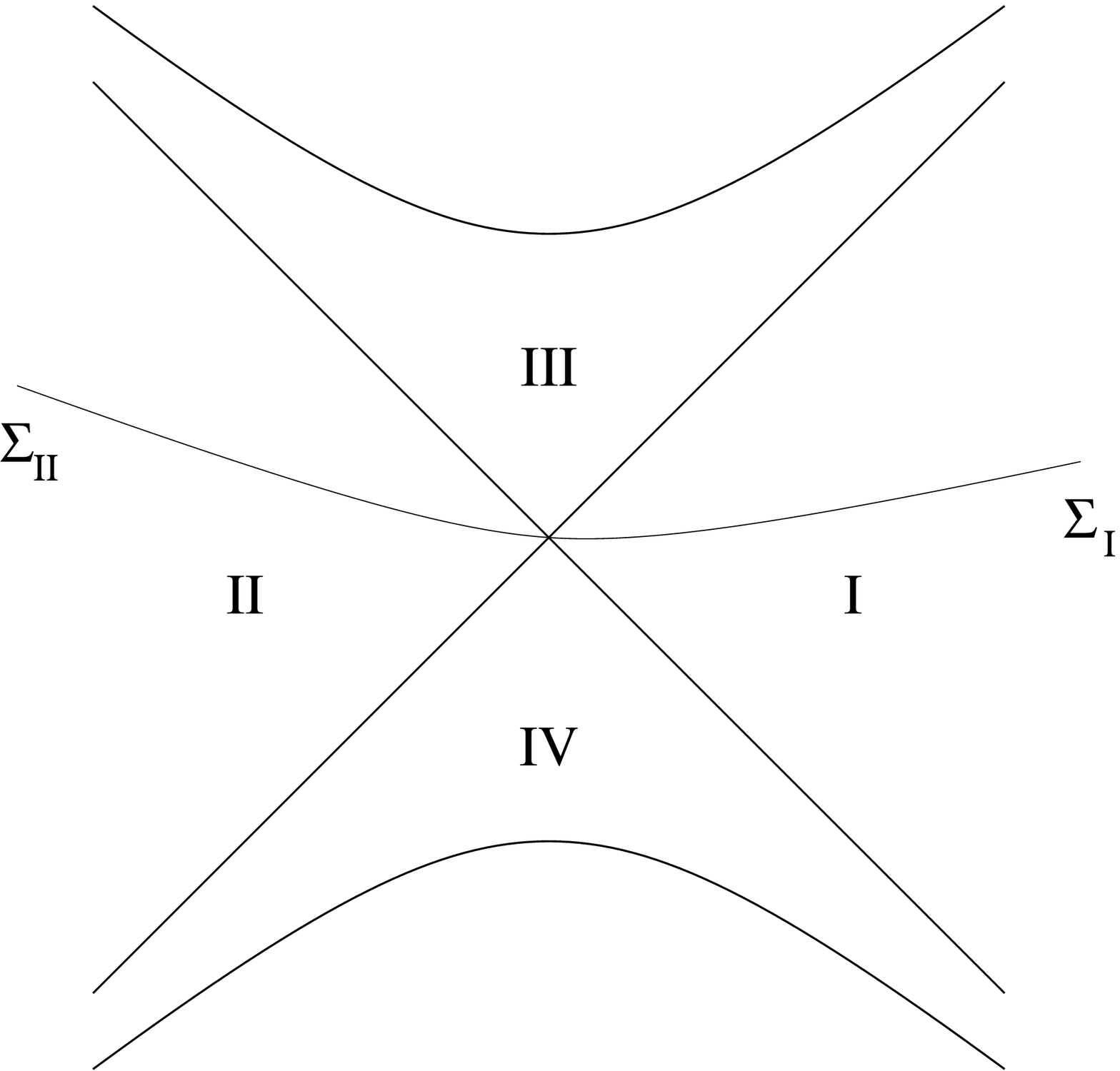,width=\anchopar}}
	\end{figure}
	\vfill
\fi


\begin{thebibliography}{99}

\bibitem{[CGHS]} C.G. Callan, S.B. Giddings, J.A. Harvey and A. 
Strominger, {\it Phys. Rev.} {\bf D} 45 R(1992)1005.  

\bibitem{[Harvey]} J.A. Harvey and A. Strominger, 
preprint EFI-92-41, hep-th/9209055 (1992).  

\bibitem{[Witten]} E. Witten, {\it Phys. Rev. } {\bf D} 44 (1991) 314. 

\bibitem{[Mandal]} G. Mandal, A. Sengupta and S. Wadia, 
{\it Mod. Phys. Lett.} {\bf A} 6 (1991) 1685.  

\bibitem{[Hirano]} S. Hirano, Y. Kazama and Y. Satoh, {\it Phys. Rev.} 
{\bf D} 48 (1993) 1687. 

\bibitem{[Verlinde]} K. Schoutens, E. Verlinde and H. Verlinde, {\it Phys. 
Rev.} {\bf D} 48 (1993) 2670. 

\bibitem{[Mikovic]} A. Mikovic, {\it Black Holes and Non-Perturbative 
Canonical 2D Dilaton Gravity}, Imperial-TP/93-94/16. 

\bibitem{[MikovicandNavarro]} A. Mikovic and M. Navarro, {\it Phys. Lett.} 
{\bf B} 315 (1993) 267-276. 

\bibitem{[NavarroandTalavera]} J. Navarro-Salas and C.F. Talavera, 
{\it Quantum Cosmological
Approach to 2d Dilaton Gravity}, preprint 
FTUV/93-34, IFIC/93-34 (to appear in {\it Nucl. Phys.} {\bf B}). 

\bibitem{[CrnkovicandWitten]} C. Crnkovi\'c and E. Witten, in {\it Three Hundred 
Years of Gravitation}, eds. S.W. Hawking and W. Israel 
(Cambridge, 1987) p. 676. 

\bibitem{[NPB]} J. Navarro-Salas, M. Navarro and V. Aldaya, {\it 
Phys. Lett.} {\bf B} 287 (1992)109; {\it Nucl. Phys.} {\bf B} 403 
(1993) 291. 

\bibitem{[Soh]} K.S. Soh, {\it Phys. Rev.} {\bf D} 49 (1994) 1906. 

\bibitem{[Thieman]} T. Thieman and H.A. Kastrup, {\it Nucl. Phys.} 
{\bf B} 399 (1993) 211. 

\bibitem{[Bilal]} A. Bilal and I.I. Kogan, 
{\it Phys. Rev.} {\bf D} 47 (1993) 5408. 

\bibitem{[Coleman]} S. Coleman, {\it Aspects of Symmetry} (Cambridge, 1985). 

\bibitem{[Ashtekaretal]} A. Ashtekar, L. Bombelli and O. Reula, in 
{\it Mechanics, Analysis and Geometry: 200 Years after Lagrange}, 
ed. M. Francavigilia (ESP, 1991) p. 417. 

\bibitem{[Crnkovic]} C. Crnkovi\'c, {\it Nucl. Phys.} 
{\bf B} 288 (1987) 419. 

\bibitem{[Birkhoff]} 
	S. W. Hawking and G. F. R. Ellis, {\it The Large Scale
Structure of Space-Time}, Cambridge University Press, Cambridge
(1973).

\bibitem{[Wittenii]}
	E. Witten, {\it Nucl. Phys.} {\bf B} 311 (1988) 46.

\bibitem{[kuchar]}
        K. Kuchar, {\it Geometrodynamics of Schwarzschild Black Holes},
preprint UU-REL-94-3-1.


\end{thebibliography}
\end{document}